%% file: paper.tex
\documentclass{article}
\usepackage{spconf}

\usepackage{graphicx}
\usepackage{amsmath,amssymb,amsfonts,bm}
\usepackage[table]{xcolor}
\usepackage{booktabs}
\usepackage{subcaption}

\usepackage{pgfplots,pgfplotstable,tikz}
\pgfplotsset{compat=1.18}
\usetikzlibrary{arrows.meta}

\usepackage{hyperref,mathtools,makecell}
\usepackage{indentfirst}
\newcommand{\img}{\bm{x}}
\newcommand{\qimg}{\hat{\img}}
\newcommand{\modes}{\bm{m}}
\newcommand{\latent}{\bm{y}}

\newcommand{\qlatent}{\hat{\latent}}

\newcommand{\analysis}{g_a}
\newcommand{\synthesis}{g_s}
\newcommand{\entropymodel}{p}
\newcommand{\quantizer}{Q}

\title{Spatial competition for low-complexity learned image compression}

\newif\ifanonymous

\ifanonymous
  \name{\mbox{}}
  \address{
    \mbox{}\\
    \mbox{}
  }
\else
  \name{Théophile Blard$^{\dagger\star}$ \qquad Pierrick Philippe$^{\dagger}$ \qquad Théo Ladune$^{\dagger}$ \qquad Xiaoran Jiang$^{\star}$ \qquad Olivier Déforges$^{\star}$}
  \address{
      $^{\dagger}$ Orange Research, France \quad \texttt{\small firstname.lastname@orange.com} \\
      $^{\star}$ Univ Rennes, INSA Rennes, CNRS, IETR (UMR 6164), France \quad \texttt{\small firstname.lastname@insa-rennes.fr}
  }
\fi

\begin{document}

\maketitle

\begin{abstract}
    Autoencoder-based image codecs achieve state-of-the-art compression performance but often incur high computational complexity, particularly at decoding time.
    This work introduces a low-complexity learned image compression framework based on spatial competition between multiple specialized neural codecs.
    For each image region, the encoder selects the codec that best matches the local content according to a rate-distortion cost.
    A mode map is transmitted as side information to indicate the per-region codec selection.
    At decoding time, this mode map-based selection guides reconstruction while preserving the complexity of a single codec.
    This design enables per-image adaptation with low decoding complexity and fast encoding.
    On the CLIC~2020 dataset, our method achieves up to $-14.5\%$ rate reduction compared to a single codec and reaches HEVC-level performance with a decoding complexity of 1433~MACs per pixel.
\end{abstract}


\input{figures/pgfplots_styles}

\section{Introduction and related works}
\label{sec:introduction}

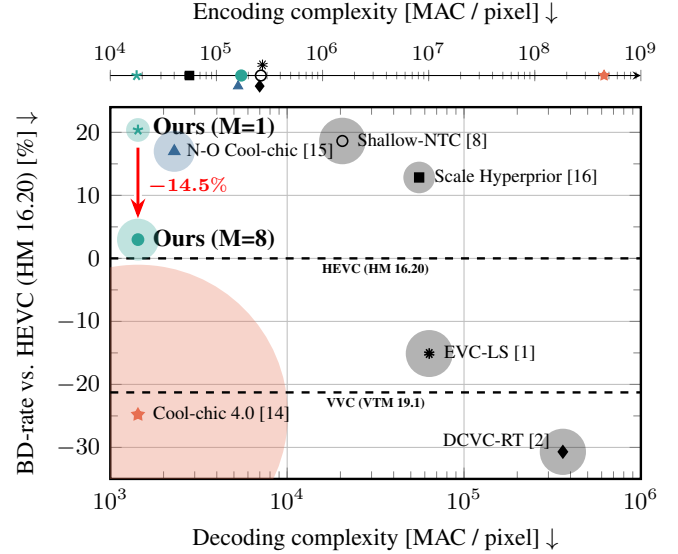
\begin{figure}[t]
    \centering
    \input{figures/decoding_complexity}
    \caption{\textbf{Rate savings vs.\ HEVC} on the CLIC~2020~\cite{toderici2020clic} dataset as a function of decoding complexity.
        Negative values mean fewer bits than HEVC at equal quality.
        \textbf{Circle area denotes encoding complexity.}
        Values are reported in Table~\ref{tab:encoder-decoder-complexity}.
    }
    \label{fig:decoding_complexity}
\end{figure}

Learned image codecs \cite{wang2023evc, jia2025towards} have recently surpassed conventional codecs (HEVC \cite{sullivan2012overview}, VVC \cite{overview-vvc-bross}) in rate-distortion (RD) performance.
These methods typically train autoencoders end-to-end to minimize the average RD cost over a large dataset.
Once trained, their parameters are fixed, and compression of unseen images relies solely on \textit{generalization}.
Achieving strong generalization typically requires complex neural networks, leading to high decoding complexity, often exceeding $10^5$ multiplications per pixel (Fig. \ref{fig:decoding_complexity}).
Such complexity might hinder practical deployment, particularly for decoding on resource-constrained devices.
\\

Reducing computational cost has therefore become an active area of research \cite{zhang2024efficient, muckley2025architecture, tan2025grouped}.
Decoding complexity is particularly critical, motivating asymmetric architectures that shift most of computation to the encoder side while keeping the decoder low in complexity \cite{yang2023computationally, wang2024asymllic}.
Beyond architectural design, model compression techniques such as structured pruning \cite{wang2023evc, johnston2019computationally, hossain2024structured} and knowledge distillation \cite{chen2025knowledge} have also proved to be effective.
Nevertheless, these approaches still incur substantially higher decoding complexity than conventional codecs and often require dedicated hardware accelerators (e.g., GPUs) to achieve practical decoding throughput.
\\

An alternative paradigm is proposed by overfitted codecs such as Cool-chic~\cite{ladune2023cool, coolchic-v4}, which jointly learn (overfit) a lightweight decoder and latent representation per image rather than aiming for generalization.
The decoder parameters are conveyed alongside the latents in the bitstream.
This yields rate-distortion performance on par with VVC with a decoder complexity on the order of a few thousand multiplications per pixel, but requires a costly iterative optimization process at the encoder.
Non-overfitted (\mbox{N-O}) Cool-chic~\cite{blard2024overfitted} replaces per-image optimization with a learned encoder that generates the latents in a single forward pass, and a universal decoder, thereby significantly accelerating encoding while preserving fast decoding.
However, this comes at the cost of reduced performance, suggesting that per-image adaptation may be particularly important when decoder capacity is limited.
\\

This work aims to reconcile low decoding complexity and fast encoding in learned image compression.
Inspired by mode selection in conventional coding, we investigate spatial competition between multiple neural codecs as an efficient mechanism for content adaptation.
The proposed method relies on a set of pretrained codecs, each specialized for specific image content, and selects at inference the most suitable codec for each region.
The selection is conveyed to the decoder via a mode map transmitted in the bitstream.
This design enables per-image adaptation without requiring the iterative optimization process used by overfitted codecs.
\\

Our contributions are summarized as follows:
\begin{itemize}
    \item \textbf{Content adaptation via spatial competition.}
          We introduce a computationally efficient content adaptation mechanism that performs rate-distortion-based spatial selection among learned image codecs.
    \item \textbf{Specialization of learned codecs.}
          An offline training framework is proposed to learn a set of neural codecs specialized for different image content.
          
    \item \textbf{Experimental evaluation.}
          The approach is evaluated on multiple datasets and compared against representative low-complexity learned codecs and conventional codecs.
\end{itemize}

\section{Method}
\label{sec:method}

\subsection{Autoencoder-based learned image compression}
\label{ssec:background-lic}

Learned image compression \cite{balle2018variational}
is commonly formulated within the transform coding paradigm using autoencoder architectures
optimized under a rate-distortion objective.
It comprises an analysis transform $\analysis(\cdot;\theta)$, a synthesis transform $\synthesis(\cdot;\phi)$ and an entropy model $\entropymodel(\cdot;\psi)$, parameterized by $\theta$, $\phi$, and $\psi$, respectively.
Given an input image $\img$, the analysis transform produces a latent representation $\latent$, which is quantized to $\qlatent$ with a quantization function $\quantizer$ and entropy coded using the entropy model.
The synthesis transform reconstructs the image $\qimg$ from the quantized latents:
\begin{equation}
    \label{eq:bg-lic-components}
    \latent = \analysis(\img;\theta), \quad
    \qlatent = \quantizer(\latent), \quad
    \qimg = \synthesis(\qlatent;\phi).
\end{equation}

The model is optimized end-to-end by minimizing the rate-distortion loss:
\begin{equation}
    \mathcal{L}(\img;\theta,\phi,\psi)
    = D(\img, \qimg) + \lambda R(\qlatent),
\end{equation}
where $D$ is a distortion metric (here, the MSE), $R$ denotes the estimated bitrate
and $\lambda$ controls the trade-off.
During training, the rate term is estimated via the entropy model as
\begin{equation}
    R(\qlatent) = - \log_2 \entropymodel(\qlatent;\psi).
\end{equation}

\subsection{Spatial competition of neural codecs}
\label{ssec:method-spatial-mode-selection}

The learned image compression framework can be extended by allowing the use of multiple codecs in competition.
Specifically, we consider a set of $M$ neural codecs,
each defined by a distinct set of parameters
$\{\theta_m, \phi_m, \psi_m\}_{m=1}^M$ for the analysis, synthesis, and entropy models.
Since the codecs are learned offline, all parameter sets are fixed after training and available to the encoder and the decoder.
\newline

\begin{figure}[t]
    \centering
    \includegraphics[width=\linewidth]{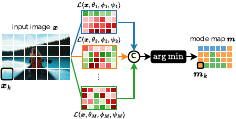}
    \caption{\textbf{Mode selection.}
        The input image is first partitioned into non-overlapping patches.
        For each patch, the $M$ codecs are evaluated by their
        rate-distortion costs, and the codec minimizing this cost is selected,
        yielding the mode map $\modes$.}
    \label{fig:mode-selection}
\end{figure}

\textbf{Mode selection.}
The input image $\img \in \mathbb{R}^{3 \times H \times W}$ is partitioned into non-overlapping $P \times P$ patches
$\{\img_k\}_{k=1}^{K}$, where $K = \lceil H/P \rceil \cdot \lceil W/P \rceil$.
For each patch $\img_k$, a mode index $\modes_k \in \{1,\dots,M\}$ is selected
to minimize the rate-distortion loss:
\begin{equation}
    \modes_k
    = \underset{m \in \{1,\dots,M\}}{\arg\min}\;
    \mathcal{L}\!\left(\img_k;\,\theta_m, \phi_m, \psi_m\right).
    \label{eq:mode-selection-indep}
\end{equation}
Fig.~\ref{fig:mode-selection} illustrates the mode selection process.
This per-patch selection ignores spatial interactions between neighboring patches
but enables mode determination in a parallel fashion.
The resulting mode map $\modes = \{\modes_k\}_{k=1}^{K}$ is encoded using fixed-length
coding and added to the bitstream.
This induces a rate overhead of
\begin{equation}
    R(\modes)
    =
    \frac{K \,\left\lceil \log_2 M \right\rceil}{H W} \; \text{bpp}.
    \label{eq:mode-selection-rate-mode-map}
\end{equation}

\begin{figure}[t]
    \centering
    \includegraphics[width=\linewidth]{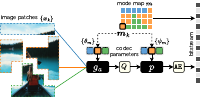}
    \caption{\textbf{Spatial encoding process.}
        Given the mode map $\modes$, each image patch is encoded using the selected analysis transform and entropy model.
        Patches sharing the same mode are processed continuously. \texttt{AE} denotes arithmetic encoding.}
    
    \label{fig:encoding-process}
\end{figure}

\textbf{Encoding.}
Given the image patches $\{\img_k\}_{k=1}^{K}$ and the mode map $\modes$,
each patch is encoded using the analysis transform specified by the selected mode $\modes_k$:
\begin{equation}
    \latent_k = \analysis(\img_k; \theta_{\modes_k}), \qquad
    \qlatent_k = \quantizer(\latent_k).
\end{equation}
As shown in Fig.~\ref{fig:encoding-process}, patches assigned to the same mode are processed jointly as connected regions, while boundaries between patches with different modes are treated as local image borders.
The resulting quantized latent patches $\{\qlatent_k\}_{k=1}^{K}$ are entropy-coded
using the selected entropy model parameters and written
to the bitstream together with the mode map $\modes$.
\newline

\textbf{Decoding.}
The mode map $\modes$ is first retrieved from the bitstream.
For each patch index $k$, the corresponding latent representation $\qlatent_k$
is entropy-decoded using the entropy model parameters $\psi_{\modes_k}$.
Each decoded latent patch is then reconstructed using the corresponding synthesis
transform:
\begin{equation}
    \qimg_k = \synthesis(\qlatent_k; \phi_{\modes_k}).
\end{equation}
As in encoding, patches sharing the same mode are reconstructed continuously.
The image patches $\{\qimg_k\}_{k=1}^{K}$ are finally assembled to form
the reconstructed image $\qimg$.
\newline

\textbf{Computational complexity.}
During mode selection, each patch is evaluated independently for the $M$ codecs,
requiring $M$ forward passes through the analysis transform, synthesis transform,
and entropy model.
After the mode map $\modes$ is determined,
each patch is processed once more through the selected analysis transform and entropy model.
Let $\kappa_{\analysis}$, $\kappa_{\synthesis}$, and $\kappa_{\entropymodel}$
denote the cost of a single forward pass through the analysis transform,
synthesis transform, and entropy model, respectively.
The total encoding complexity is therefore given by
\begin{equation}
    \label{eq:encoding-complexity}
    \kappa_{\text{enc}}
    =
    \underbrace{
        M\left(
        \kappa_{\analysis}
        + \kappa_{\synthesis}
        + \kappa_{\entropymodel}
        \right)
    }_{\text{mode selection}}
    \;+\;
    \underbrace{
        \kappa_{\analysis} + \kappa_{\entropymodel}
    }_{\text{final encoding}}.
\end{equation}

The decoding complexity remains independent of $M$, as each patch requires a single pass in the selected entropy model and synthesis transform:
\begin{equation}
    \label{eq:decoding-complexity}
    \kappa_{\text{dec}}
    = \kappa_{\synthesis} + \kappa_{\entropymodel}.
\end{equation}

\begin{figure*}[t]
    \centering
    \input{figures/qualitative.tex}
    \caption{\textbf{Visualization of mode maps for different numbers of modes.}
        Mode maps are obtained with different numbers of competing modes~($M$) at $\lambda=0.0002$.
        Images are \textit{stefan-kunze-26931} and \textit{nomao-saeki-33553} from the CLIC~2020 dataset.}
    \label{fig:qualitative}
\end{figure*}
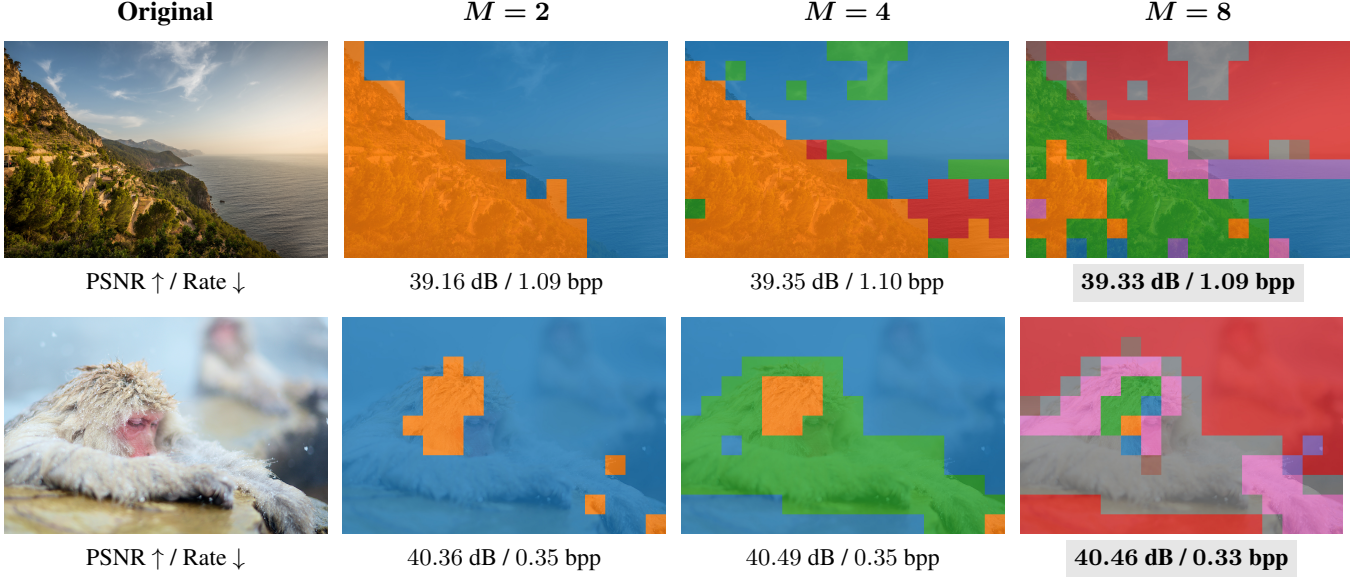

\subsection{Training specialized codecs for competition}
\label{sec:method-codec-specialization}

The proposed spatial competition method relies on the availability of multiple specialized codecs.
Such specialization requires finding an appropriate repartition of the training data, so that each codec is adapted to a subset of image patches.
Given a dataset of image patches $\mathcal{D} = \{\bm{x}_i\}_{i=1}^N$,
our goal is to jointly learn a set of codec parameters
$\{\theta_m, \phi_m, \psi_m\}_{m=1}^M$ and a cluster assignment
$z_i \in \{1,\dots,M\}$ for each training patch, such that the overall rate-distortion cost is minimized:
\begin{equation}
    \label{eq:method-rd-cost}
    \min_{\{z_i\}, \{\theta_m, \phi_m, \psi_m\}}
    \sum_{i=1}^{N}
    \mathcal{L}\!\left(\bm{x}_i;\,\theta_{z_i}, \phi_{z_i}, \psi_{z_i}\right).
\end{equation}

Similar to clustering and codebook-learning methods in vector quantization~\cite{gray1984vector} and transform competition~\cite{arrufatbatalla2015multiple}, this optimization is performed by alternating between an \emph{assignment step} and an \emph{update step}.
\newline

\textbf{Assignment step.}
Each training sample $\img_i$ is assigned to the codec that yields the lowest rate-distortion loss:
\begin{equation}
    \label{eq:method-cluster-assign}
    z_i \leftarrow
    \underset{m \in \{1,\dots,M\}}{\arg\min}\;
    \mathcal{L}\!\left(\img_i;\,\theta_m, \phi_m, \psi_m\right).
\end{equation}
This induces a partition of the training set into $M$ clusters,
\begin{equation}
    \label{eq:method-cluster-partitioning}
    \mathcal{D}_m = \{\img_i \mid z_i = m\}.
\end{equation}

\textbf{Update step.}
Given the current assignments, the parameters of each codec are updated by minimizing the rate-distortion loss over the samples assigned to its cluster:
\begin{equation}
    \label{eq:method-cluster-update}
    \theta_m, \phi_m, \psi_m \leftarrow \underset{\theta,\phi,\psi}{\arg\min} \sum_{\img_i \in \mathcal{D}_m} \mathcal{L}(\bm{x}_i, \theta, \phi, \psi)
\end{equation}
Unlike vector quantization or transform competition methods, where updates are derived analytically, the update step here is performed via gradient-based optimization of the neural network parameters.
This procedure alternates between assignment and update steps until the cluster assignments converge.

\section{Experiments}
\label{sec:experiments}

\subsection{Experimental setup}
\label{ssec:experiments-experimental-setup}

\textbf{Model architecture.}
All experiments are conducted using the same autoencoder architecture.
To enable low-complexity decoding, the synthesis transform and entropy model are taken from Cool-chic~4.0~\cite{coolchic-v4}, with respective complexities of $\kappa_{\synthesis}=708$ and $\kappa_{\entropymodel}=725$~MAC/pixel.
To form a complete autoencoder, a compatible analysis transform derived from~\cite{blard2024overfitted} is adopted.
Compared to the original architecture, the number of channels and residual blocks is reduced to control encoding complexity, resulting in $\kappa_{\analysis}=18$~kMAC/pixel.
\newline

\textbf{Training.}
Models and cluster assignments are optimized through repeated \textit{assignment} and \textit{update} steps.
At initialization, training samples are randomly assigned such that all clusters have the same size.
Each \textit{update} step consists of $100$k training iterations, with the learning rate annealed to zero using a cosine schedule.
The initial learning rate is set to $10^{-3}$ and is progressively reduced between phases by a multiplicative factor $\alpha=0.98$.
Training samples consist of randomly cropped $128\times128$ patches from the Unsplash~\cite{chesser2020unsplash} dataset.
Models are trained using the Adam \cite{kingma2014adam} optimizer with a batch size of $64$.
Backpropagation through quantization is enabled using additive noise and softround relaxation~\cite{kim2024c3}.
Separate runs are conducted for each rate constraint $\lambda \in \{0.01, 0.004, 0.001, 0.0004, 0.0002, 0.0001\}$ and each number of clusters, with up to $M=8$.
\newline

{
    \setlength{\tabcolsep}{3pt}
    \begin{table}
        \centering
        \input{tables/bdrate_vs_m1.tex}
        
        \caption{
            \textbf{BD-rate savings relative to a single codec} for different numbers of modes $M$.
            Encoding complexity \(\kappa_{\mathrm{enc}}\) is reported in kMAC/pixel.
        }
        \label{tab:bdrate_vs_m1}
    \end{table}
}

\textbf{Evaluation.}
We evaluate on the Kodak~\cite{kodak}, CLIC~2020~\cite{toderici2020clic} and JPEG~AI Test~\cite{jpegai2020test} datasets, covering image resolutions from $512\times384$ to $3680\times2456$ pixels.
Consistently with the training setup, test-time inference uses a patch size of $P=128$ and up to $M=8$ competing codecs.
Rate-distortion performance is measured using PSNR on RGB channels and codecs are compared using the Bjøntegaard Delta Rate~\cite{bjontegaard2001calculation} (BD-Rate).
Comparisons are made against HEVC~\cite{sullivan2012overview} (HM~16.20) and several low-complexity learned codecs, including Cool-chic~4.0~\cite{coolchic-v4}, N-O~Cool-chic~\cite{blard2024overfitted}, Scale Hyperprior~\cite{balle2018variational}, Shallow-NTC~\cite{yang2023computationally}, and EVC~\cite{wang2023evc}.

\subsection{Results}
\label{ssec:experiments-results}

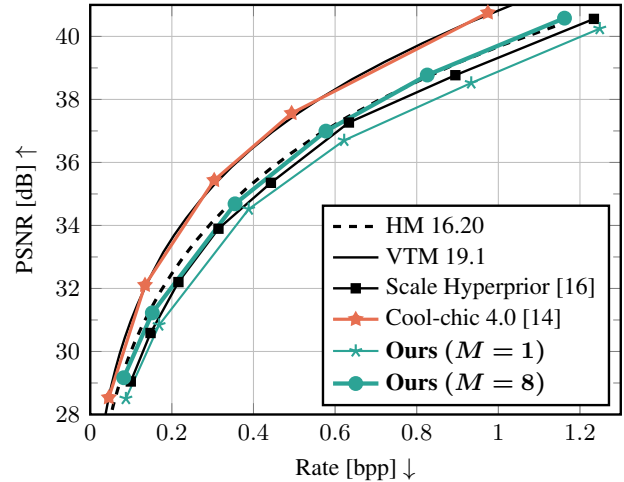
\begin{figure}[t]
    \centering
    \input{figures/rd_curve_psnr}
    \caption{\textbf{Rate-distortion performance} on CLIC~2020.}
    \label{fig:rd_curve_psnr}
\end{figure}

\textbf{Rate-distortion performance.}
Table~\ref{tab:bdrate_vs_m1} shows that increasing the number of modes $M$  consistently improves compression efficiency.
For $M=8$, BD-rate savings reach up to $-14.5\%$ on the CLIC~2020 dataset compared to the single-codec baseline ($M=1$), demonstrating the effectiveness of spatial competition.
Fig.~\ref{fig:qualitative} further illustrates that larger values of $M$ improve rate-distortion performance.
The gains are modest on Kodak, which can be attributed to the smaller image resolution: the limited number of spatial patches limits the exploitation of spatial mode selection.
Fig.~\ref{fig:rd_curve_psnr} reports rate-distortion curves, where consistent rate savings are observed across the entire bitrate range.
Depending on the number of competing codecs, the encoding complexity ranges from 18 to 171~kMAC/pixel, which remains within the typical range of standard autoencoder-based methods.
Importantly, increasing $M$ does not affect decoding complexity, which remains fixed at $\kappa_{\mathrm{dec}}=1433$~MAC/pixel.
\newline

\begin{figure}[t]
    \centering
    \input{figures/encoding_complexity}
    \caption{
        \textbf{BD-rate vs. encoding complexity} on CLIC~2020.
        Following \cite{blard2024overfitted}, encoding complexity for Cool-chic 4.0 is computed as $3 \times \kappa_{\mathrm{dec}} \times$ number of iterations.
    }
    \label{fig:bdrate-vs-encoding}
\end{figure}
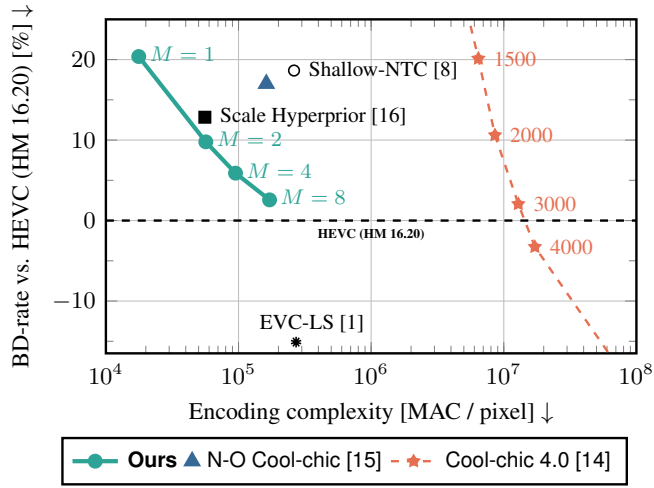

\textbf{Comparison to prior work.}
Fig.~\ref{fig:bdrate-vs-encoding} and Table~\ref{tab:encoder-decoder-complexity} position compression performance against encoding and decoding complexity.
Results are reported relative to HEVC (HM~16.20) on the CLIC~2020 dataset.
With $M=2$ codecs, our approach outperforms Scale Hyperprior~\cite{balle2018variational}, while using a decoder that is $40\times$ less complex (1.4 vs.\ 56~kMAC/pixel).
With $M=8$, our method reaches near-HEVC performance ($+2.6\%$ BD-rate).
At comparable encoding complexity, it significantly outperforms Shallow-NTC~\cite{yang2023computationally}, which remains at $+18.6\%$ BD-rate vs HEVC despite a $15\times$ more complex decoder (1.4 vs.\ 21~kMAC/pixel), and improves upon N-O Cool-chic~\cite{blard2024overfitted} by $11.8\%$ BD-rate. 
With identical decoding complexity, overfitted Cool-chic~\cite{coolchic-v4} achieves superior compression performance; but at the cost of a substantially heavier encoding process.
\newline

\begin{table}
    \centering
    \renewcommand{\arraystretch}{0.9}
    \input{tables/comparison_prior_work.tex}
    \caption{\textbf{BD-rate vs.\ HEVC} on CLIC~2020. Encoding and decoding complexity $\kappa_{enc}$ and $\kappa_{dec}$ are reported in kMAC/pixel.}
    \label{tab:encoder-decoder-complexity}
\end{table}

\textbf{Mode-map overhead.}
Signaling the mode map incurs a negligible side-information cost.
With $M=8$ the mode map requires only $3$ bits per patch, corresponding to $1.8\times10^{-4}$~bpp for patch size $P=128$.
Across all tested datasets and operating points, this represents at most $1\%$ of the total rate.
This low overhead leaves room to increase the number of modes $M$ or to reduce the patch size $P$ for finer adaptation.
\newline

\textbf{Decoder memory footprint.}
The number of decoder parameters scales linearly with the number of modes $M$, as each mode is associated with its own synthesis transform and entropy model.
However, these components are extremely compact (on the order of a thousand parameters), so the resulting memory overhead remains well below that of a single state-of-the-art autoencoder-based codec.

\section{Conclusion}
\label{sec:conclusion}

This work introduces a low-complexity learned image compression framework based on spatial competition between multiple specialized neural codecs.
By selecting, for each image region, the codec that minimizes the local rate-distortion cost, the proposed approach enables efficient per-image adaptation.
On the CLIC~2020 dataset, our method achieves up to $-14.5\%$ rate reduction relative to a single-codec configuration and delivers rate-distortion results comparable to HEVC, while maintaining a low decoding complexity of 1433~MACs per pixel.
Importantly, the encoding complexity remains in the typical range of autoencoder-based codecs; in contrast to overfitted codecs, no costly iterative optimization is required.
These results demonstrate that spatial competition can be an effective mechanism for improving the performance of low-complexity learned codecs.
\\

\textbf{Limitations and future work.}
While the proposed approach demonstrates an attractive trade-off between complexity and performance, it still lags behind state-of-the-art methods in absolute rate-distortion performance.
Future work will therefore focus on narrowing this gap.
Promising directions include increasing the number of competing codecs and reducing the patch size to enable finer spatial adaptation.
Exposing codecs to multiple patch sizes during training and inference may further improve performance by better capturing content at different scales.
Finally, extending the spatial competition framework to other learned compression architectures would help validate its broader applicability.

\vfill\pagebreak


{
    \small
    \bibliographystyle{IEEEbib}
    \bibliography{refs}
}

\end{document}

%% file: figures/pgfplots_styles.tex
\definecolor{myblue}{RGB}{55,105,150}
\definecolor{myred}{RGB}{231,111,81}
\definecolor{mygreen}{RGB}{44,163,149}

\pgfdeclareplotmark{sharp star}{
    \pgfpathmoveto{\pgfpointpolar{90}{2pt}} 
    \foreach \i in {1,...,5} {
            \pgfpathlineto{\pgfpointpolar{90 + \i*144 - 72}{1pt}} 
            \pgfpathlineto{\pgfpointpolar{90 + \i*144}{2pt}}     
        }
    \pgfpathclose
    \pgfusepathqfillstroke
}

\pgfplotsset{
    myaxis/.style={
            width=\linewidth,
            height=6cm,
            grid=major,
            minor y tick num=1,
            ytick distance=10,
            legend columns=3,
            legend style={
                    font=\footnotesize\sffamily,
                    at={(0.45,-0.25)},
                    anchor=north,
                    draw=black
                },
            line width=0.9pt,
            tick label style={font=\small},
            label style={font=\small},
            title style={font=\footnotesize, yshift=-1ex},
            legend cell align=left
        },
    series/.style={
            mark size=2pt,
            nodes near coords,
            point meta=explicit symbolic,
            every node near coord/.append style={font=\footnotesize}
        },
    hyper/.style={
            mygreen,
            ultra thick,
            mark=*,
            mark size=2pt,
            every node near coord/.append style={
                    anchor=west,
                    yshift=1pt
                }
        },
    ours/.style={
            mygreen,
            ultra thick,
            mark=*,
            mark size=2pt,
            every node near coord/.append style={
                    anchor=west,
                    yshift=1pt
                }
        },
    nocool/.style={
            myblue,
            dashed,
            mark=triangle*,
            mark size=2pt,
            mark options={solid},
            every node near coord/.append style={
                    anchor=south,
                    yshift=2pt
                }
        },
    cool/.style={
            myred,
            dashed,
            mark=sharp star,
            mark options={solid},
            every node near coord/.append style={
                    anchor=west,
                    xshift=2pt
                }
        },
    hm/.style={
            black,
            dashed,
            very thick
        },
    refpoint/.style={
            mark=square*,
            mark size=2pt,
            black,
            font=\footnotesize,
            forget plot
        },
    proportionalhalo/.style={
            only marks,
            mark=*,
            mark options={draw=none},
            fill opacity=0.3,
            draw opacity=0,
            line width=0pt
        },
}

%% file: figures/decoding_complexity.tex
\newcommand{\halosizefromenc}[2]{%
  \pgfkeys{/pgf/fpu=true}%
  \pgfmathparse{sqrt((#2^0.5)*0.15)}%
  \pgfmathfloatparsenumber{\pgfmathresult}%
  \pgfmathfloattofixed{\pgfmathresult}%
  \edef#1{\pgfmathresult}%
  \pgfkeys{/pgf/fpu=false}%
}

\halosizefromenc{\haloNoCool}{160000}
\halosizefromenc{\haloCool}{4.50e08}
\halosizefromenc{\haloOursSingle}{17690}
\halosizefromenc{\haloOurs}{171399}
\halosizefromenc{\haloShallow}{262150}

\halosizefromenc{\haloEvc}{271735}
\halosizefromenc{\haloScaleHP}{55750}
\halosizefromenc{\haloAsymllic}{208800}
\halosizefromenc{\haloDCVC}{255500}

\begin{tikzpicture}

  \begin{semilogxaxis}[
      myaxis, name=main,
      width=\linewidth,
      height=6.5cm,
      xlabel={Decoding complexity [MAC / pixel] $\downarrow$},
      ylabel={BD-rate vs. HEVC (HM 16.20) [\%] $\downarrow$},
      ymin=-35, ymax=24, xmin=1e3, xmax=1e6,
      clip=true
    ]

    \addplot[
      proportionalhalo,
      fill=myblue,
      mark size=\haloNoCool
    ]
    table [x=k_dec, y=bdrate]
      {data/complexity/no-cool-chic.tsv};

    \addplot[nocool]
    table [x=k_dec, y=bdrate]
      {data/complexity/no-cool-chic.tsv}
    node[
        anchor=west,
        xshift=1pt,
        text=black,
        font=\scriptsize
      ]{N-O Cool-chic~\cite{blard2024overfitted}};

    \addplot[proportionalhalo, fill=myred,  mark size=\haloCool, clip marker paths=true]
    coordinates {(1433, -24.787)};
    \addplot[cool, mark=sharp star,] coordinates {(1433, -24.787)}
    node[anchor=west, xshift=2pt, text=black, font=\scriptsize]{Cool-chic 4.0~\cite{coolchic-v4}};

    \addplot[proportionalhalo, fill=mygreen, mark size=\haloOursSingle]
    coordinates {(1433, 20.38)};
    \addplot[hyper, thick, mark=star] coordinates {(1433, 20.38)}
    node[anchor=west, xshift=2pt, yshift=1pt, text=black, font=\small\bfseries]{Ours (M=1)};

    \addplot[proportionalhalo, fill=mygreen, mark size=\haloOurs]
    coordinates {(1433, 2.98)};
    \addplot[hyper, thick] coordinates {(1433, 2.98)}
    node[anchor=west, xshift=2pt, text=black,  font=\small\bfseries]{Ours (M=8)};

    \addplot[proportionalhalo, mark size=\haloScaleHP]
    coordinates {(55750, 12.84)};
    \addplot[refpoint, mark=square*, text=black, mark size=1.5] coordinates {(55750, 12.84)}
    node[anchor=west, xshift=2pt, font=\scriptsize]{Scale Hyperprior~\cite{balle2018variational}};

    \addplot[proportionalhalo,  mark size=\haloShallow, clip marker paths=true]
    coordinates {(20520, 18.63)};
    \addplot[refpoint, mark=o, semithick, text=black, ] coordinates {(20520, 18.63)}
    node[anchor=west, xshift=2pt, font=\scriptsize]{Shallow-NTC~\cite{yang2023computationally}};

    \addplot[proportionalhalo, mark size=\haloEvc]
    coordinates {(63573, -15.1)};
    \addplot[refpoint, mark=10-pointed star, semithick, text=black] coordinates {(63573, -15.1)}
    node[anchor=west, xshift=2pt,font=\scriptsize]{EVC-LS~\cite{wang2023evc}};

    \addplot[proportionalhalo, fill=black, mark size=\haloDCVC, clip marker paths=true]
    coordinates {(362200, -30.698)};
    \addplot[refpoint, mark=diamond*] coordinates {(362200, -30.698)}
    node[anchor=east, xshift=-1pt, yshift=4pt, font=\scriptsize]{DCVC-RT~\cite{jia2025towards}}; 

    \addplot[black, dashed, domain=1e3:1e6]{0};
    \node[
      anchor=south,
      font=\tiny\bfseries,
      text=black,
      text opacity=1,
      inner sep=2pt,
      rounded corners=4pt
    ] at (axis description cs:0.5,0.53) {HEVC (HM 16.20)};

    \addplot[black, dashed, domain=1e3:1e6]{-21.27};
    \node[
      anchor=south,
      font=\tiny\bfseries,
      text=black,
      text opacity=1,
      inner sep=2pt,
      rounded corners=4pt
    ] at (axis description cs:0.5,0.17) {VVC (VTM 19.1)};

    \draw[-{Stealth[length=2.6mm]}, line width=1.1pt, red,
    shorten >=8pt, shorten <=6pt,
    preaction={draw=white, line width=3pt}]
    (axis cs:1433,20.10) -- (axis cs:1433,2.98)
    node[midway, right, font=\scriptsize\bfseries, text=red] {$\bm{-14.5\%}$};

  \end{semilogxaxis}

  \begin{semilogxaxis}[
      at={(main.north)}, anchor=south,   
      width=\linewidth, height=2.0cm,
      xmin=1e4, xmax=1e9,
      ymin=0.9, ymax=1.1,
      axis x line=top,
      axis y line=none,
      ytick=\empty, ymin=0, ymax=1,
      xtick={1e4, 1e5,1e6,1e7,1e8,1e9},
      xticklabel style={font=\footnotesize},
      xlabel={Encoding complexity [MAC / pixel] $\downarrow$},
      xlabel style={font=\small},
      enlargelimits=false        
    ]

    \addplot[hyper, thick, fill=mygreen, mark=star, mark size=2pt]
    coordinates {(17690,1)};

    \addplot[hyper, semithick, fill=mygreen, mark size=2pt]
    coordinates {(171399,1)};

    \addplot[nocool, only marks, fill=myblue,mark=triangle*, mark size=2pt, yshift=-4pt]
    coordinates {(160000,1)};

    \addplot[refpoint, mark=square*,semithick, mark size=1.5pt]
    coordinates {(55750,1)};

    \addplot[refpoint, mark=diamond*,semithick, mark size=2pt, yshift=-4pt]
    coordinates {(255500,1)};

    \addplot[refpoint, mark=10-pointed star, mark size=2pt,  yshift=4pt]
    coordinates {(271735,1)};

    \addplot[refpoint,  mark=o, semithick, semithick, mark size=2pt]
    coordinates {(262150,1)};

    \addplot[cool,  mark=sharp star,  fill=myred, mark size=2pt]
    coordinates {(4.5e8,1)};
  \end{semilogxaxis}

\end{tikzpicture}

%% file: figures/qualitative.tex
\makebox[0.24\linewidth][c]{\textbf{Original}}
\hfill
\makebox[0.24\linewidth][c]{\bm{$M=2$}}
\hfill
\makebox[0.24\linewidth][c]{\bm{$M=4$}}
\hfill
\makebox[0.24\linewidth][c]{$\bm{M=8}$}

\vspace{0.5em}

\begin{minipage}[t]{0.24\linewidth}
    \centering
    \includegraphics[width=\linewidth]{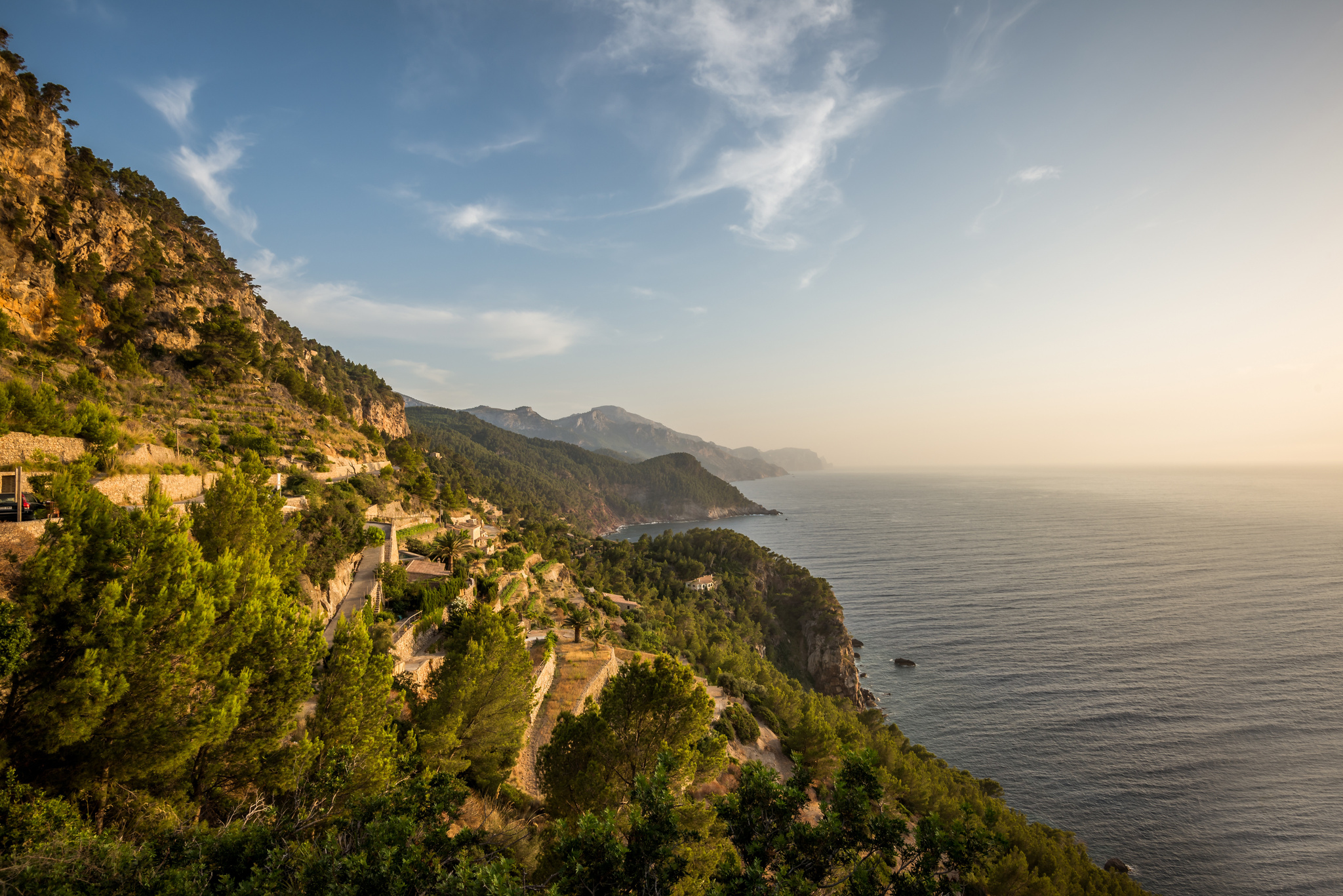}
    \vspace{1pt}
    {\small PSNR $\uparrow$ / Rate $\downarrow$}
\end{minipage}\hfill
\begin{minipage}[t]{0.24\linewidth}
    \centering
    \includegraphics[width=\linewidth]{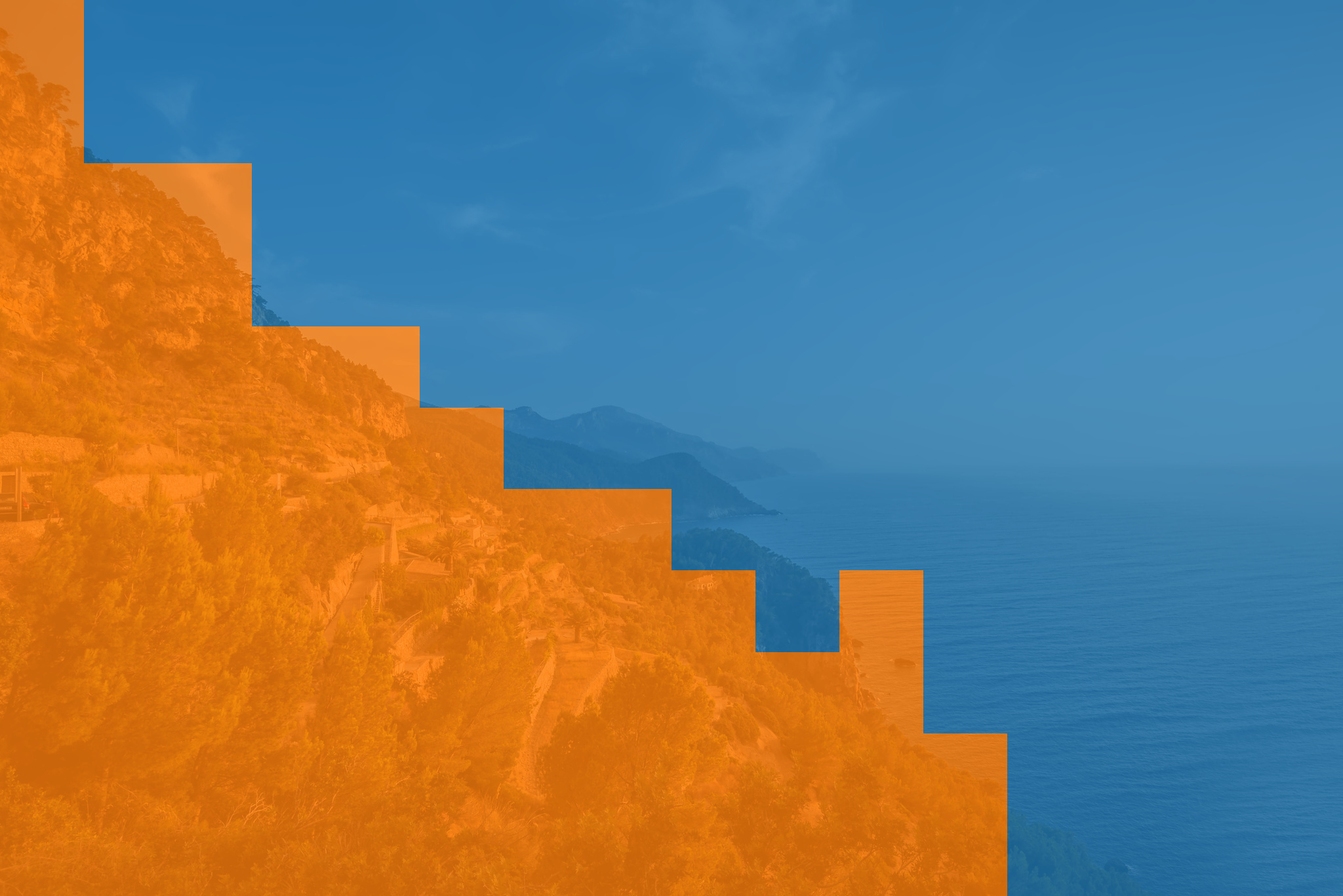}
    \vspace{1pt}
    {\small $39.16$ dB / $1.09$ bpp}
\end{minipage}\hfill
\begin{minipage}[t]{0.24\linewidth}
    \centering
    \includegraphics[width=\linewidth]{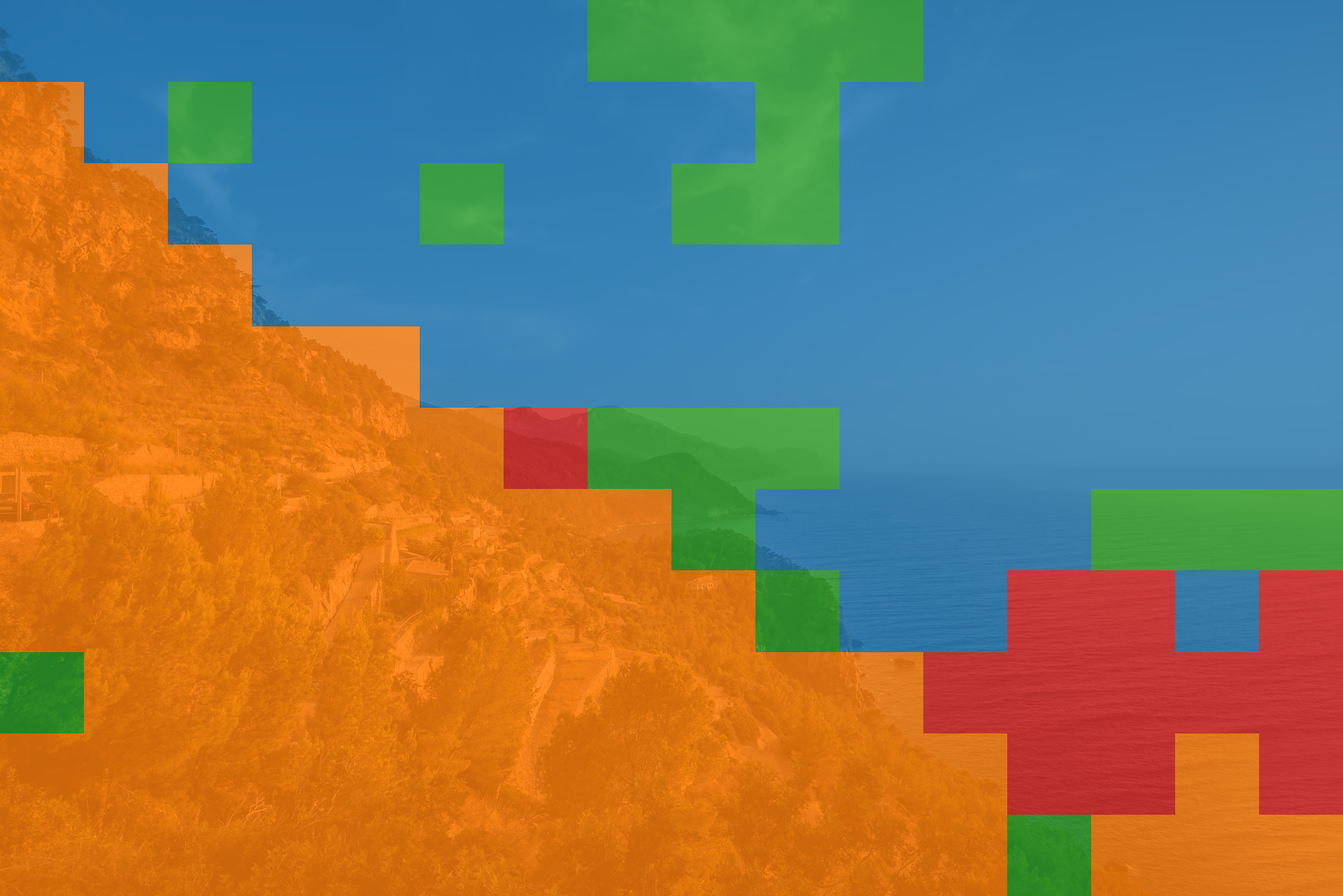}
    {\small $39.35$ dB / $1.10$ bpp}
\end{minipage}\hfill
\begin{minipage}[t]{0.24\linewidth}
    \centering
    \includegraphics[width=\linewidth]{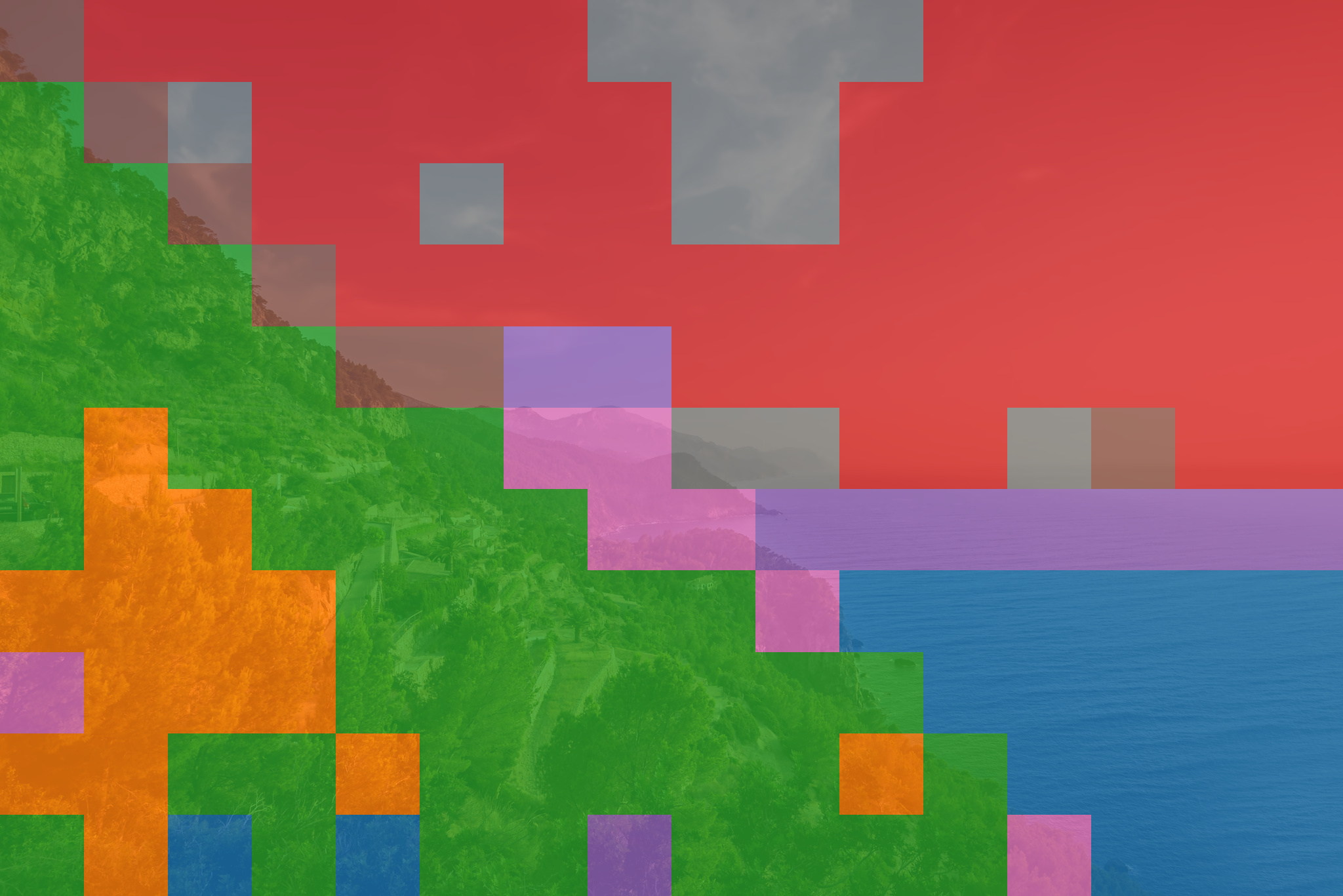}
    {\small \colorbox{gray!20}{$\bm{39.33}$ \textbf{dB /} $\bm{1.09}$ \textbf{bpp}}}
\end{minipage}

\vspace{0.5em}

\begin{minipage}[t]{0.24\linewidth}
    \centering
    \includegraphics[width=\linewidth]{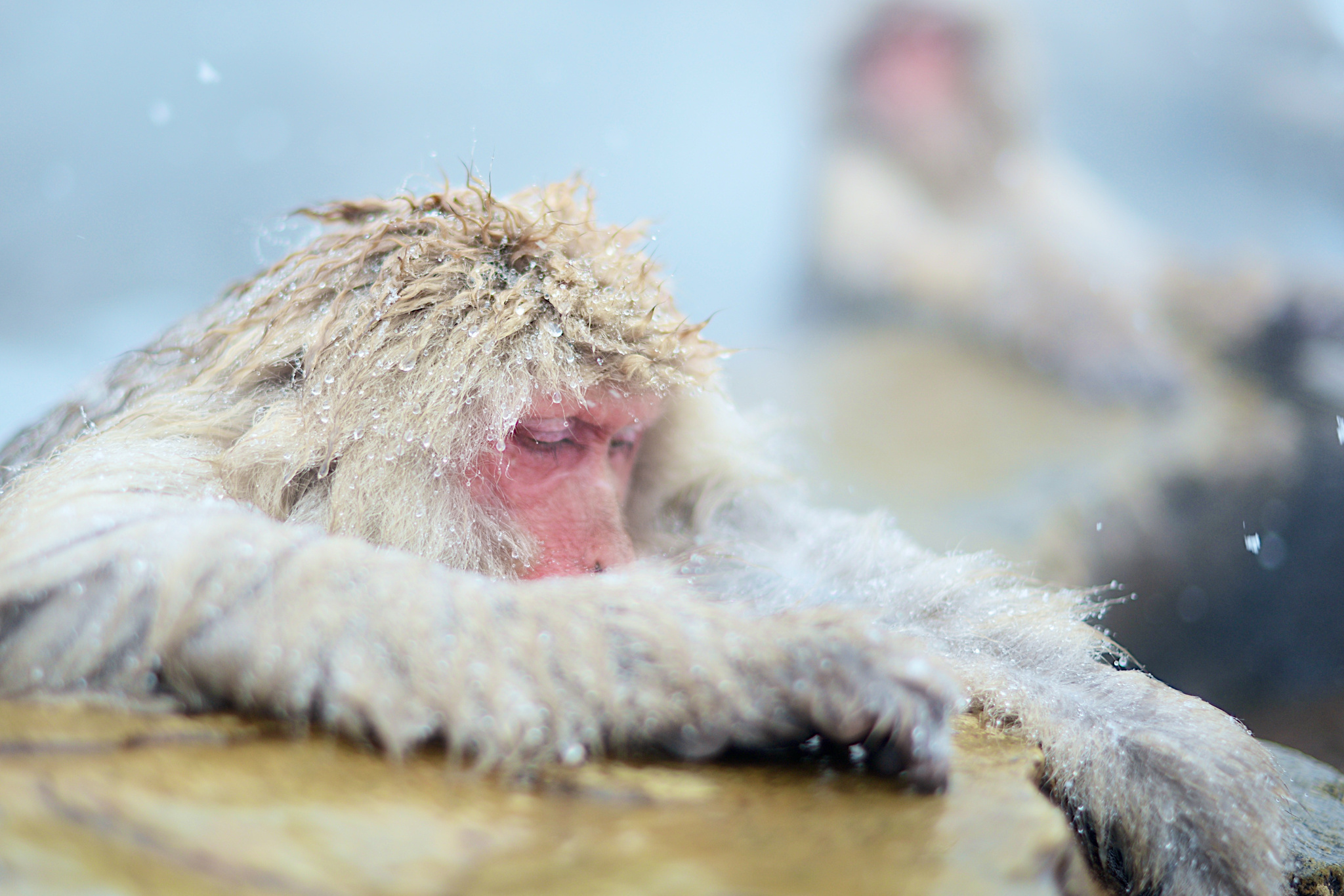}
    \vspace{1pt}
    {\small PSNR $\uparrow$ / Rate $\downarrow$}
\end{minipage}\hfill
\begin{minipage}[t]{0.24\linewidth}
    \centering
    \includegraphics[width=\linewidth]{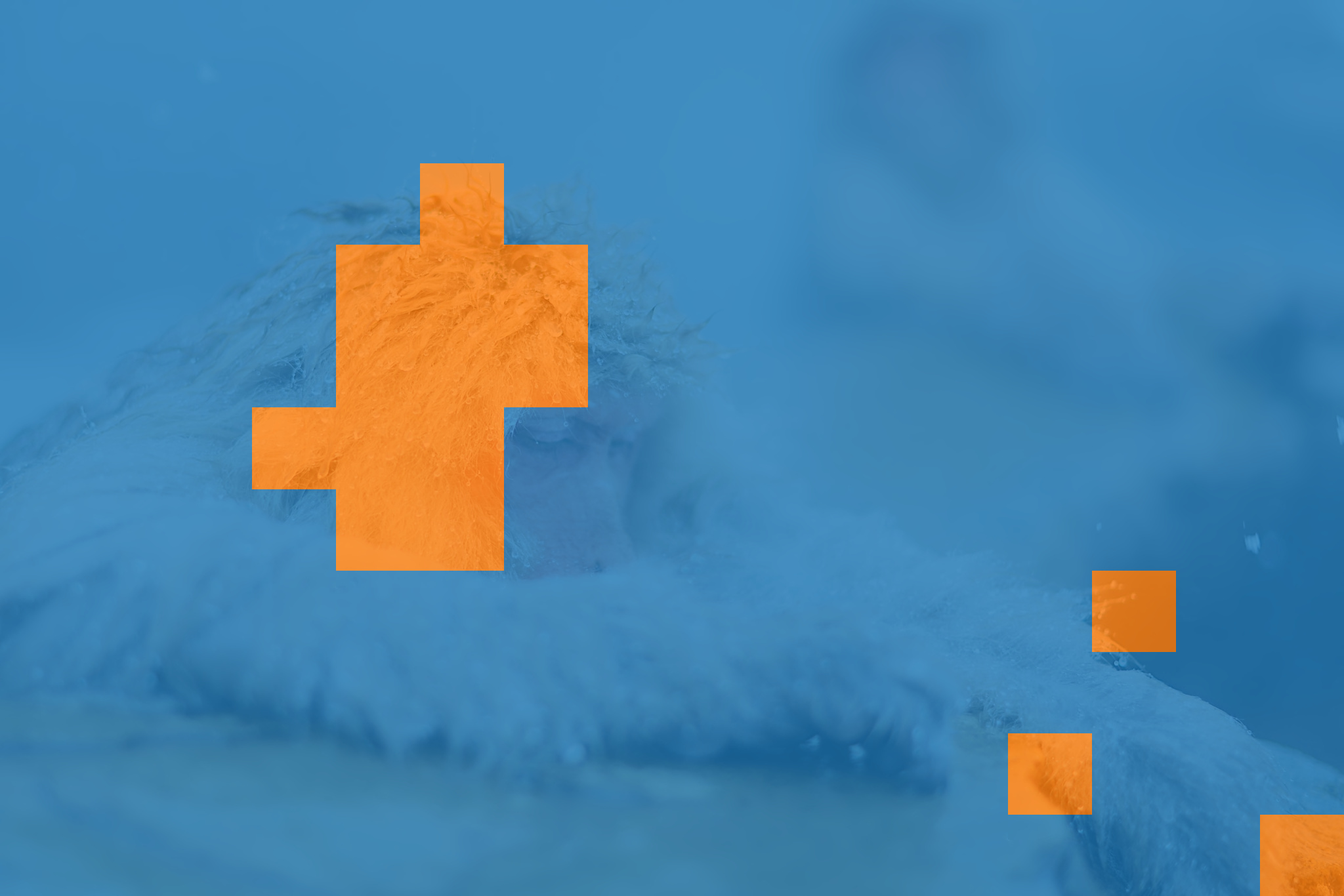}
    \vspace{1pt}
    {\small $40.36$ dB / $0.35$ bpp}
\end{minipage}\hfill
\begin{minipage}[t]{0.24\linewidth}
    \centering
    \includegraphics[width=\linewidth]{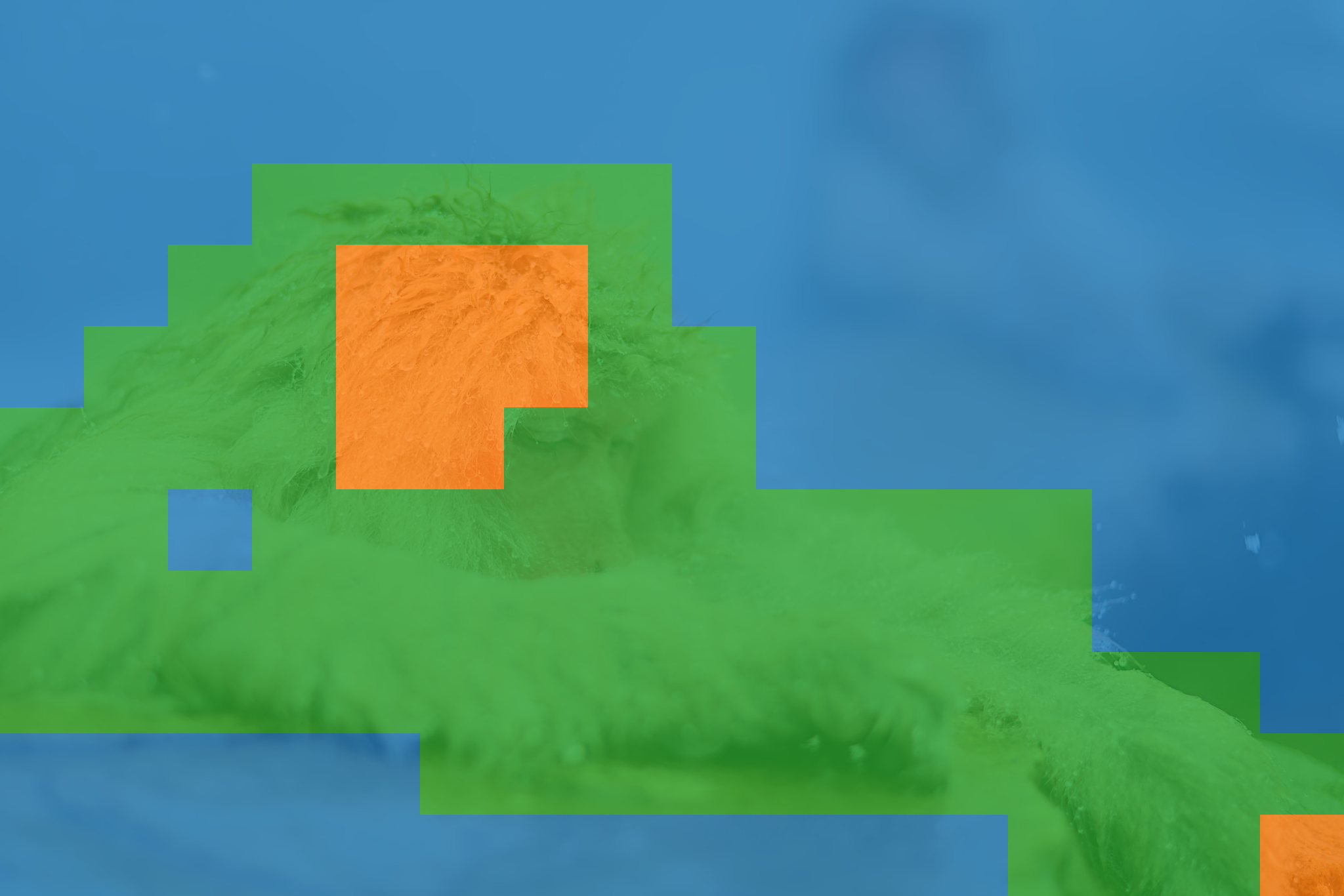}
    \vspace{1pt}
    {\small $40.49$ dB / $0.35$ bpp}
\end{minipage}\hfill
\begin{minipage}[t]{0.24\linewidth}
    \centering
    \includegraphics[width=\linewidth]{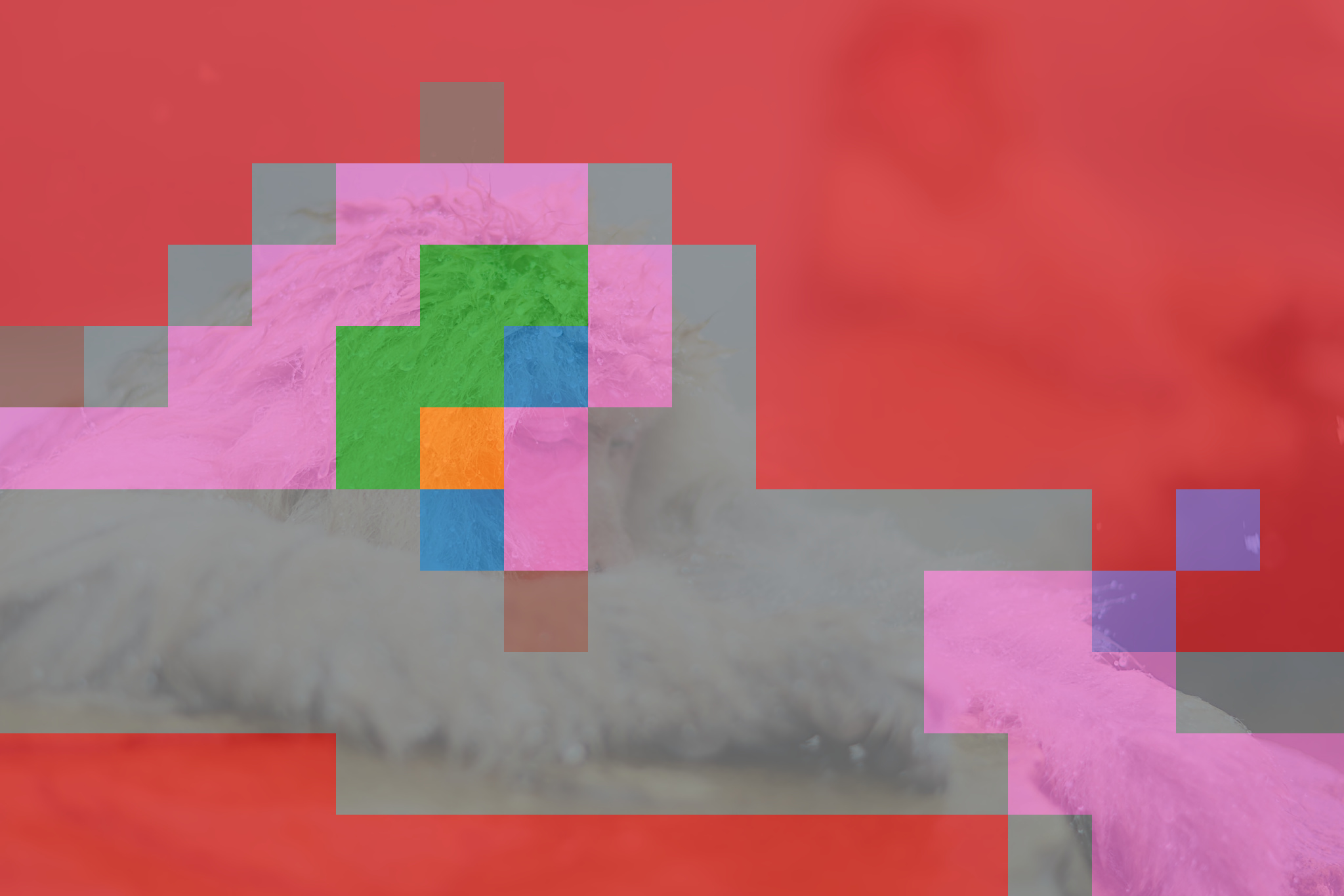}
    {\small \colorbox{gray!20}{$\bm{40.46}$ \textbf{dB /} $\bm{0.33}$ \textbf{bpp}}}
\end{minipage}

%% file: tables/bdrate_vs_m1.tex
\begin{tabular}{l c c c c}
    \toprule
             &                                        & \multicolumn{3}{c}{BD-rate (\%) vs. single codec ($M=1$) $\downarrow$}                                                                     \\
    \cmidrule(lr){3-5}
    $M$      & ${\kappa_{\mathrm{enc}}}$ $\downarrow$ & Kodak~\cite{kodak}                                                     & CLIC~2020~\cite{toderici2020clic} & JPEG~AI~\cite{jpegai2020test} \\
    \midrule
    $1$      & $18$                                   & $0$                                                                    & $0$                               & $0$                           \\

    $2$      & $57$                                   & $-4.1$                                                                 & $-8.6$                            & $-6.2$                        \\
    $4$      & $95$                                   & $-5.6$                                                                 & $-11.8$                           & $-8.7$                        \\
    \rowcolor{gray!20}
    $\bm{8}$ & $171$                                  & $\bm{-7.4}$                                                            & $\bm{-14.5}$                      & $\bm{-11.4}$                  \\
    \bottomrule
\end{tabular}



%% file: figures/encoding_complexity.tex
\begin{tikzpicture}
  \begin{semilogxaxis}[
      myaxis,
      xlabel={Encoding complexity [MAC / pixel] $\downarrow$},
      ylabel={BD-rate vs. HEVC (HM 16.20) [\%] $\downarrow$},
      xmin=1e4, xmax=1e8,
      ymin=-16.5, ymax=25,
      legend style={row sep=4pt, nodes={inner sep=3pt}}
    ]

    \addplot[series, ours]
    table [x=k_enc, y=bdrate, meta index=0]
      {data/complexity/ours.tsv};
    \addlegendentry{\textbf{Ours}}

    \addplot[series, nocool, mark size=3pt, only marks]
    table [x=k_enc, y=bdrate]
      {data/complexity/no-cool-chic.tsv};
    \addlegendentry{N-O Cool-chic~\cite{blard2024overfitted}}

    \addplot[series, cool, mark size=3pt]
    table [x=k_enc, y=bdrate, meta index=0]
      {data/complexity/coolchic.tsv};
    \addlegendentry{Cool-chic 4.0~\cite{coolchic-v4}}

    \addplot[refpoint, mark=o, semithick, black]
    table [x=k_enc, y=bdrate]
      {data/complexity/shallow-ntc.tsv}
    node[anchor=west, xshift=2pt]
      {Shallow-NTC~\cite{yang2023computationally}};

    \addplot[refpoint, mark=10-pointed star, semithick, black]
    table [x=k_enc, y=bdrate]
      {data/complexity/evc-ls.tsv}
    node[anchor=south, xshift=6pt]
      {EVC-LS~\cite{wang2023evc}};

    \addplot[refpoint, mark=square*, black]
    table [x=k_enc, y=bdrate]
      {data/complexity/scale-hyperprior.tsv}
    node[anchor=west, xshift=2pt]
      {Scale Hyperprior~\cite{balle2018variational}};

    \addplot[black, dashed, domain=1e4:1e9]{0};
    \node[
      anchor=south,
      font=\tiny\bfseries,
      inner sep=2pt,
      rounded corners=4pt
    ] at (axis description cs:0.5,0.32)
    {HEVC (HM 16.20)};

  \end{semilogxaxis}
\end{tikzpicture}

%% file: figures/rd_curve_psnr.tex
\begin{tikzpicture}
  \begin{axis}[
      myaxis,
      xlabel={Rate [bpp] $\downarrow$},
      ylabel={PSNR [dB] $\uparrow$},
      height=7cm,
      xmin=0, xmax=1.3,
      ymin=28, ymax=41,
      ytick={24,26,28,30,32,34,36,38,40,42,44},
      legend columns=1,
      legend style={
          at={(0.98,0.02)},
          anchor=south east,
          draw=black,
          fill=white,
          font=\small,
          legend cell align={left}
        }
    ]

    \addplot[series, hm]
    table [x=rate_bpp, y=psnr_db, col sep=tab]
      {data/clic20-pro-valid/hm_average.tsv};
    \addlegendentry{HM 16.20}

    \addplot[series]
    table [x=rate_bpp, y=psnr_db, col sep=tab]
      {data/clic20-pro-valid/vtm_average.tsv};
    \addlegendentry{VTM 19.1}

    \addplot[series, mark=square*, solid, mark size=1.5pt]
    table [x=rate_bpp, y=psnr_db, col sep=tab]
      {data/clic20-pro-valid/balle18_average.tsv};
    \addlegendentry{Scale Hyperprior~\cite{balle2018variational}}

    \addplot[series, cool, solid, very  thick, mark=sharp star]
    table [x=rate_bpp, y=psnr_db, col sep=tab]
      {data/clic20-pro-valid/cool-chic-v4_average.tsv};
    \addlegendentry{Cool-chic 4.0 \cite{coolchic-v4}}

    \addplot[series, hyper, thick, mark=star, mark size=2.5pt]
    table [x=rate_bpp, y=psnr_db, col sep=tab]
      {data/clic20-pro-valid/ours-m1-average.tsv};
    \addlegendentry{\textbf{Ours $\bm{(M=1)}$}}

    \addplot[series, hyper, solid, ultra  thick, ]
    table [x=rate_bpp, y=psnr_db, col sep=tab]
      {data/clic20-pro-valid/ours-m8-average.tsv};
    \addlegendentry{\textbf{Ours $\bm{(M=8)}$}}

  \end{axis}
\end{tikzpicture}

%% file: tables/comparison_prior_work.tex
\newcommand{\tkmark}[3][]{%
  \tikz[baseline=-0.6ex,scale=#2]{%
    \if\relax\detokenize{#1}\relax
    \else\color{#1}\fi
    \pgfuseplotmark{#3}%
  }%
}

\newcommand{\tksharpstar}{\tkmark{1}{sharp star}}
\newcommand{\tkstar}{\tkmark{1}{star}}
\newcommand{\tksquare}{\tkmark{1}{square*}}
\newcommand{\tkdiamond}{\tkmark{1}{diamond*}}
\newcommand{\tktriangle}{\tkmark{1}{triangle*}}
\newcommand{\tkcircle}{\tkmark{1}{o}}
\newcommand{\tkpointedstar}{\tkmark{1}{10-pointed star}}

\newcommand{\tkbluetriangle}{\tkmark[myblue]{1}{triangle*}}
\newcommand{\tkbluecircle}{\tkmark[myblue]{0.7}{*}}
\newcommand{\tkredcircle}{\tkmark[myred]{0.7}{*}}
\newcommand{\tkgreencircle}{\tkmark[mygreen]{1}{*}}
\newcommand{\tkgreenstar}{\tkmark[mygreen]{1}{star}}
\newcommand{\tkredsharpstar}{\tkmark[myred]{1}{sharp star}}

\newcommand{\gr}{\rowcolor[gray]{0.95}}

\begin{tabular}{lccc}
  Method                                                     & $\kappa_{enc}$ $\downarrow$ & $\kappa_{dec}$ $\downarrow$ & BD-rate [\%] $\downarrow$ \\
  \Xhline{1.0pt}
  \tkdiamond{} DCVC-RT~\cite{jia2025towards}                 & 256                         & 362                         & $-30.7$                   \\
  \tkredsharpstar{} Cool-chic 4.0~\cite{coolchic-v4}         & $\sim10^6$                  & 1.4                         & $-24.8$                   \\
  \tkpointedstar{} EVC-LS~\cite{wang2023evc}                 & 272                         & 64                          & $-15.1$                   \\

  \tksquare{} Scale Hyperprior~\cite{balle2018variational}   & 56                          & 56                          & $+12.8$                   \\

  \tkbluetriangle{} N-O Cool-chic~\cite{blard2024overfitted} & 162                         & 2.3                         & $+17.1$                   \\
  \tkcircle{} Shallow-NTC~\cite{yang2023computationally}     & 262                         & 21                          & $+18.6$                   \\

  \Xhline{1.0pt}
  \gr
  \tkgreenstar{} \textbf{Ours ($\bm{M=1}$)}                  & $\bm{18}$                   & \textbf{1.4}                & $\bm{+20.4}$              \\
  \gr
  \tkgreencircle{} \textbf{Ours ($\bm{M=2}$)}                & $\bm{57}$                   & \textbf{1.4}                & $\bm{+9.8}$               \\
  \gr
  \tkgreencircle{} \textbf{Ours ($\bm{M=4}$)}                & $\bm{95}$                   & \textbf{1.4}                & $\bm{+5.9}$               \\
  \gr
  \tkgreencircle{} \textbf{Ours ($\bm{M=8}$)}                & $\bm{171}$                  & \textbf{1.4}                & $\bm{+2.6}$               \\
  \Xhline{1.0pt}
\end{tabular}